\documentclass[final,sort&compress,3p,times]{elsarticle}
\usepackage{multirow}
\usepackage{array}
\usepackage[numbers]{natbib}

\usepackage{amssymb}
\usepackage{url}            % simple URL typesetting
\usepackage{booktabs}       % professional-quality tables
\usepackage{amsfonts}       % blackboard math symbols
\usepackage{nicefrac}       % compact symbols for 1/2, etc.
\usepackage{microtype}      % microtypography
\usepackage{lipsum}
\usepackage{fancyhdr}       % header
\usepackage{graphicx}       % graphics
\graphicspath{{media/}}     % organize your images and other figures under media/ folder

\journal{Journal Name}

\begin{document}

\begin{frontmatter}



\title{Searching for Strange Quark Matter Objects Among White Dwarfs}


\author[1,2,3]{Abdusattar Kurban\corref{cor1}} 
\ead{lompa46@163.com}

\author[2,4]{Yong-Feng Huang\corref{cor1}}
\ead{hyf@nju.edu.cn}

\author[5] {Jin-Jun Geng}

\author[6,7,8,9]{Hong-Shi Zong}

\address[1]{Xinjiang Astronomical Observatory, Chinese Academy of Sciences, Urumqi 830011, Xinjiang, People's Republic of China}

\address[2]{School of Astronomy and Space Science, Nanjing University, Nanjing 210023, Jiangsu, People's Republic of China}

\address[3]{Xinjiang Key Laboratory of Radio Astrophysics, Urumqi 830011, Xinjiang, People's Republic of China}	

\address[4]{Key Laboratory of Modern Astronomy and Astrophysics (Nanjing University), Ministry of Education,  Nanjing 210023, Jiangsu, People's Republic of China}

\address[5]{Purple Mountain Observatory, Chinese Academy of Sciences, Nanjing 210023, Jiangsu, People's Republic of China}

\address[6]{Department of Physics, Nanjing University, Nanjing 210093, Jiangsu, People's Republic of China}

\address[7]{Joint Center for Particle, Nuclear Physics and Cosmology, Nanjing 210093, Jiangsu, People's Republic of China}

\address[8]{Nanjing Institute of Proton Source Technology, Nanjing 210046, Jiangsu, People's Republic of China}

\address[9]{Department of Physics, Anhui Normal University, Wuhu 241000, Anhui, People's Republic of China}

\cortext[cor1]{Corresponding author}

\begin{abstract}
The ground state of matter may be strange quark
matter (SQM), not hadronic matter. A whole sequence of SQM objects, ranging
from strange quark stars and strange quark dwarfs to strange quark planets,
can stably exist according to this SQM hypothesis.
A strange dwarf has a mass similar to that of a normal white dwarf, but could
harbor an extremely dense SQM core (with the density as large as
$\sim \rm 4\times10^{14}\,g\,cm^{-3} $) at the center so that its radius can be
correspondingly smaller. In this study, we try to search for strange dwarfs
among the observed ``white dwarfs'' by considering their difference in the
mass-radius relation. Seven strange dwarf candidates are identified in this
way, whose masses are in the range of $\sim 0.02$ --  $0.12 M_{\odot}$,
with the radii narrowly distributed in $\sim$ 9,000 -- 15,000 km.
The seven objects are LSPM J0815+1633, LP 240-30,
BD+20 5125B, LP 462-12, WD J1257+5428, 2MASS J13453297+4200437,
and SDSS J085557.46+053524.5.
Comparing with white dwarfs of similar mass, these candidates are obviously smaller
in radius. Further observations with large radio/infrared/optical telescopes on these
interesting candidates are solicited.
\end{abstract}


\end{frontmatter}

%%\linenumbers


% Main text
\section{Introduction}\label{Introduction}

It has long been argued that the ground state of matter may be quark
matter \citep{Itoh1970,Bodmer1971,Witten1984} rather than hadronic form, because
its energy per baryon could be less than that of the most stable atomic
nucleus ($^{56}\mathrm{Fe}$).
The existence of more exotic states such as strange quark matter (SQM -- an
approximately equal mixture of u, d, s quarks) in the core of compact
stars was also speculated \citep{Itoh1970,Bodmer1971,Witten1984,Farhi1984}.
According to this SQM hypothesis, there could exist a whole sequence of SQM objects,
ranging from strange stars (SSs), with masses similar to those of neutron stars (NSs)
\citep{Witten1984,Farhi1984,Alcock1986,Glendenning1992,Kettner1995PhRvD}, to strange dwarfs (SDs) -- stellar
objects composed of a small SQM core and a thick normal matter crust which could be
regarded as the counterpart of normal white dwarfs (WDs) \citep{Glendenning1995,Glendenning1995+,Weber1997+},
and even to strange quark planets \citep{Glendenning1995,Glendenning1995+,Xu2003,Horvath2012,Geng2015,Huang2017}.

According to previous studies \citep{Alcock1986,Glendenning1992,Glendenning1995,Glendenning1995+},
SQM stars may be bare SQM objects, but they also may be covered by a crust of normal
hadronic matter. From this point of view, crusted SQM objects share many common
features with ordinary compact objects.
The similarity between SQM stars and normal NSs makes it difficult to
discriminate these two internally different compact stars observationally \citep{Alcock1986}.
Anyway, a great effort has been made to try to reveal the difference between them.
For example, they may have different mass-radius ($ M-R $) relations,
cooling rates, maximum masses, gravitational wave patterns, etc. It was even suggested
that SSs can host very close-in planets so that pulsar planets with an
orbit period less than $\sim 6100$ s can be safely regarded as an indication of
the existence of SQM planetary systems \citep{Huang2017,Kuerban2020}.

There are also some potential problems associated with the existence of SSs.
First, the coexistence of SSs and NSs in our Galaxy was questioned by Madsen \citep{Madsen1988}
that if SSs exist then NSs cannot also exist due to the pollution by strangelets produced
during the merger of two SSs. However, the results of general relativistic simulations of SS
mergers \citep{Bauswein2009PhRvL} show that the flux of strangelets is very sensitive to the bag
constant of the MIT bag model. When the bag constant is large enough, the flux of
strangelets may even disappear so that other NSs would not be polluted by strangelets and would
not be converted to SSs. In other words, it is still possible for NSs and SSs to coexist in the Galaxy.
Many other studies \citep{Drago2014PhRvD,Wiktorowicz2017ApJ,Drago2018ApJ,Char2019AIPC,Pietri2019ApJ,Drago2020PhRvD}
have been conducted based on the coexistence of both SSs and NSs.
The second issue concerns the oscillation of pulsars. \citet{Watts2007MNRAS} claimed that magnetar
oscillations are not compatible with objects not having a normal crust, i.e. bare SSs,
but fit perfectly well with NSs. The key point to solving this
problem is to consider the possibility that SSs are not bare but are covered by a normal matter
crust. \citet{Mannarelli2015ApJ} studied the oscillations of non-bare SSs and showed that
the breaking of the ionic crust might generate some phenomena such as gamma-ray bursts
and quasi-periodic oscillations.
Thirdly, it has been argued that SSs may cool too rapidly to be compatible with observations \citep{Pizzochero1991}.
But note that the cooling rate of SSs could be affected by many complicated factors, such as the
uncertainties in the strange quark mass and the quark gap
energy, the effect of normal matter crust, the uncertain rates
of the quark direct/modified Urca processes and bremsstrahlung (see the review
of \citet{Weber2005} and references therein). At the same time, the measuring of
the ages and surface temperatures of pulsars is also troubled
by large uncertainties in observations. As a result, it is
still too early to draw a firm conclusion in this aspect \citep{Li2020JHEAp}.
In short, we see that more studies are still needed to understand the internal
composition of dense compact stars.

Recent discovery of several $ \sim 2 \, M_\odot $ pulsars
\citep{Demorest2010,Antoniadis2013,Cromartie2020} greatly
stimulated the study of the internal composition of these
enigmatic compact stars. \citet{Annala2020} proposed some evidence
for the existence of quark-matter cores in such massive NSs.
Especially, an interesting compact object of $ 2.6 \, M_{\odot} $
was reported to be associated with the gravitational wave event
GW190814 \citep{Abbott2020}.
Some authors argued that this massive object could be an SS
\citep{Tangphati2022EPJC,Chu2021PhRvC,Miao2021ApJ,Zhang2021PhRvD,Bombaci2021PhRvL,Roupas2021PhRvD}
or a rapidly rotating NS with an exotic SQM core
\citep{Dexheimer2021PhRvC,Rather2021PhRvC}. This further
attracts researchers' interests of trying to find SQM in compact
stars and related explosive phenomenons. One interesting idea is
that the collapse of the crust of SSs may produce FRBs, which will
serve as indirect evidence for the SQM hypothesis
\citep{ZhangY2018,Geng2021}.

Here we will mainly focus on strange dwarfs (SDs). The normal
matter crust of strange dwarfs makes them similar to normal white
dwarfs (WDs), but they still have different features in the $ M-R
$ relation. Following \citet{Glendenning1995, Glendenning1995+},
many other authors have investigated the properties of SDs and
tried to differentiate them from WDs in both theoretical
\citep{Benvenuto1996, Benvenuto1996ApJ, Weber2005,
	Marranghello2006, Benvenuto2005,
	Benvenuto2006a,Benvenuto2006b,Alford2012,Alford2017ApJ} and observational
\citep{Mathews2006, Mathews2010, Vartanyan2004, Fontaine2007, Vartanyan2014, Jiang2018} aspects.
However, note that while \citet{Glendenning1995, Glendenning1995+} suggested that
strange dwarfs are stable with respect to radial oscillation
due to the action of the density discontinuity between the SQM core and the
crust, \citet{Alford2017ApJ} recently questioned the stability of these
objects after numerically solving the Sturm-Liouville equations for
the lowest-energy modes of the star. They argued that strange dwarfs are
not stable.
In Alford et al.'s calculations, the bottom density of
the crust is typically taken as $\rm 4.3 \times10^{11} \rm \, g\, cm^{-3} $,
which corresponds to the largest possible density for
the crust, i.e. the neutron drip density.
Note that a detailed
study on the mechanical equilibrium of the crust suggests that the actual
density at the bottom of the crust should be much less \citep{Huang1997a, Huang1997b},
i.e. no larger than $ \sim \rho_{\rm drip}/5 \approx 8.3\times10^{10}
\rm \, g\, cm^{-3} $. Such a thinner crust might be
beneficial to the stability.
Additionally, other factors such as the
equation of state (EOS) of strange quark matter can also affect the stability of SDs.
For example, \citet{Jim2019PhRvD} recently investigated the effect of strange quark
mass and the interaction between quarks on the effect of EOS
by means of the perturbative quantum chromodynamics (pQCD)
method. Using their new EOS, they calculated the radial
oscillations of quark stars. It is found that the oscillation properties are very
different from that calculated under the MIT bag model.
Including the effects of the strange quark mass and quark interactions can lower the period
of the fundamental oscillation mode. Considering all these
factors, the stability of strange dwarfs may still need further studies.
Anyway, in this study, we will assume that strange dwarfs can
stably exist, and will try to search for them among white dwarfs.

Identifying strange dwarfs observationally is an important method
to reveal the enigma of these interesting objects. Mathews et al.
\citep{Mathews2006, Mathews2010} identified seven objects as
candidates of strange dwarfs because their measured masses and
radii are consistent with the expected mass-radius relation of
strange quark dwarfs with a carbon crust. The progress in
observational technology leads to a drastic increase in the number
of white dwarfs being detected in the past decades. These vast
amounts of data inspire us to re-examine all of them to identify
more strange dwarf candidates. In this study, we will study the
observed parameters of the dwarf objects systematically and try to
identify possible strange dwarfs by comparing the observational
data with theoretical modeling.

%--------------------------------------------------------------------
\section{Structure of strange dwarfs} \label{sec:model}

The internal structure of compact objects can be calculated by
solving the general relativistic form of
Tolman-Oppenheimer-Volkoff (TOV) equation
\citep{Tolman1939,Oppenheimer1939},
\begin{equation}
	\frac{dp(r)}{dr} = \frac{-G[\rho(r) + p(r)/c^2][m(r) + 4 \pi r^3 p(r)/c^2]}{r^2[1-2Gm(r)/rc^2]},
\end{equation}
\begin{equation}
	\frac{dm(r)}{dr} = 4 \pi r^{2} \rho(r),
\end{equation}
where $ m(r) $ is the mass within the radial coordinate $ r $,
$ \rho (r) $ and $ p(r) $ are mass density and pressure at $ r $, respectively;
$ G $ is the gravitational constant, and $c$ is the speed of light.
The relation between $ p(r) $ and $ \rho (r) $ is determined by the EOS which
itself depends on the internal composition of the compact object.

As described in ref.\citep{Glendenning1995,
	Glendenning1995+}, a strange dwarf is composed of SQM in the core
and normal matter in the crust. There is a strong electric field
between the core and the crust, which makes it possible that the
crust is suspended out of contact with the SQM core
\citep{Alcock1986, Kettner1995PhRvD}. The EOS of such a structure
is characterized by a discontinuity in density between the SQM and
the normal crust matter across the electric gap where the pressure
at the bottom of the nuclear crust equals the pressure at the
surface of the SQM core
\citep{Glendenning1995,Glendenning1995+,Weber2005}. The SQM core
and the normal matter crust are two distinct regions so that each
region requires a different EOS. Consequently, the property of the
strange dwarf is determined by two parameters, the central density
of the core and the density at the crust bottom ($ \rho_{\rm cb}
$).

For SQM in the core, we employe an EOS derived from the
simple MIT bag model \citep{Farhi1984,Alcock1986}, i.e. $ p(r) =
(\rho(r) c^{2} - 4B)/3 $, where $ B $ is the so called bag
constant. In our study, we take a typical value of $ B = 57 \,\rm
MeV \, fm^{-3}$. For normal matter in the crust where the density
ranges in a very wide range, we adopt the EOS proposed by Baym,
Pethick \& Sutherland (usually called as the BPS EOS)
\citep{Baym1971}. At the surface of the crust, the density is only
$\sim 8$ g cm$^{-3}$. It increases quickly when the depth
increases, but $^{56}\mathrm{Fe}$ keeps to be the dominant nuclide
as long as the density is below $\sim 8 \times 10^6$ g cm$^{-3}$.
When $8 \times 10^6$ g cm$^{-3} < \rho < 1 \times 10^9$ g
cm$^{-3}$, the dominant nuclide becomes
$^{62}\mathrm{Ni}$/$^{64}\mathrm{Ni}$. As the density further
increases, the environment becomes more and more neutron-rich, and
a series of heavy nuclide ($^{84}\mathrm{Se}$, $^{82}\mathrm{Ge}$,
$^{80}\mathrm{Zn}$, $^{78}\mathrm{Ni}$, $^{76}\mathrm{Fe}$,
$^{124}\mathrm{Mo}$, $^{122}\mathrm{Zr}$ ) will appear in
succession. Finally, at the bottom of the curst, the nuclide will
be dominated by $^{118}\mathrm{Kr}$ and the density will be up to
$4.3 \times 10^{11}$ g cm$^{-3}$. This density is called the neutron
drip density ($\rho_{\rm drip}$), above which neutrons will drip
out from nuclei. $\rho_{\rm drip}$ is believed to be the largest
possible density for $ \rho_{\rm cb} $ because free neutrons could
not be supported by the electric field, but will fall directly
onto the SQM core and be converted to quarks
\citep{Glendenning1992, Glendenning1995, Weber2005}. However, note
that a further study by \cite{Huang1997a, Huang1997b} indicates
that the maximum density at the crust bottom actually should be
significantly smaller, i.e. no larger than $ \sim \rho_{\rm
	drip}/5 \approx 8.3\times10^{10}\rm \, g\, cm^{-3} $.

Note that the BPS EOS covers a wide range of density. It
essentially can satisfactorily describe the $p$-$\rho$ relation of
matter as long as $\rho \leq \rho_{\rm drip}$ \citep{Baym1971,
	Chamel2008LRR}. It incorporates the various electron-capture and
fusion reactions associated with the appearing of different
nuclide, which allow matter composed of nuclei to reach the
cold-catalyzed ground state.

Having specified the EOS, we can solve the TOV equation by using
the Runge-Kutta method to derive the properties of SQM objects
with a normal matter crust. Our calculations will be carried out
by assuming different initial values for $ \rho_{\rm cb} $ ($
\rho_{\rm cb} = 4.3 \times 10^{11} \rm \, g\, cm^{-3} $, $ 8.3
\times 10^{10} \rm \, g\, cm^{-3}$, $ 10^{9} \rm \, g\, cm^{-3}$,
$ 10^{8} \rm \, g\, cm^{-3}$, etc) to investigate the effect of
crusts with various thicknesses. For each $ \rho_{\rm cb} $ value,
the $ M_{\rm core} $ will further vary from $ 10^{-12} \, M_\odot
$ to $ 0.1 \, M_\odot $. The results of our calculations will be
presented in Section \ref{sec:selection}.

\section{Data Collection}\label{sec:data}

The fundamental parameters of white dwarfs can be
determined through various observations, especially via the
photometric technique \citep{Blouin2019}. Here we are mainly
concerned about the mass ($M$) and radius ($R$) of white dwarfs.
We briefly describe how these parameters are measured
observationally. Firstly, the distance ($D$) can be known from
parallax measurements such as those provided by Gaia
\citep{Fusillo2019MNRAS}. Secondly, the solid angle
($\pi R^2/D^2$) and the surface temperature ($T_{\rm eff}$) can be
obtained by fitting the observed flux and the observed continuum
spectrum. From these quantities, people can easily solve for the
radius of the white dwarf. Thirdly, the surface gravity ($g$) can
also be determined by fitting the copious line features in the
spectrum, since the strength and width of the lines are sensitive
to the density of particles in the atmosphere, which is controlled
by the surface gravity
\citep{Fusillo2019MNRAS, Fontaine2001PASP}.
Note that the central wavelength of each spectrum line is also
affected by the gravitational redshift, which again reflects the
effect of the surface gravity and can be measured
\citep{Chandra2020ApJ}. Finally, with $R$
and $g$ determined above, we can easily derive the mass ($M$) of
the white dwarf from $g = G M / R^2$.

Mass and radius are important parameters that can be effectively
used to probe the intrinsic composition of stars. Especially, the
mass-radius relations should be different for strange dwarfs and
normal white dwarfs. To search for possible strange dwarf
candidates, we will systematically examine all the white dwarfs
listed in the popular Montreal White Dwarf Database (hereafter,
MWDD
\footnote{\url{http://www.montrealwhitedwarfdatabase.org/tables-and-charts.html}}
)
\citep{Dufour2017}. All together, there are about 55900 objects
listed as white dwarf in MWDD. Among them, 39041 are available
with the parameters of surface gravity ($g$) and mass ($M$).
Note that although the mass $M$ (instead of $R$) is listed
in the database for clarity, the directly measured parameter
actually is the radius $R$, as described just above. Anyway, from
the two available parameters of $g$ and $M$, we can easily
calculate the stellar radius as $ R = \left({G M}/{g}\right)^{1/2}$.
In this study, we will examine the masses and radii of these
39041 objects to search for possible strange dwarfs.

\section{Identifying strange dwarf candidates}\label{sec:selection}

\subsection{Mass-Radius Relation}\label{subsec:m-r}

Following the procedure described in Section~\ref{sec:model}, we have calculated the
theoretical $M-R$ relations for both strange dwarfs and white dwarfs.
The results are plotted in figure~\ref{fig:fig1}.
In this figure, the solid curves illustrate the mass-radius relations of
white dwarfs, under various approximation.
The solid magenta curve represents the $M-R$ relation in Chandrasekhar's
model \citep{Chandrasekhar1967}. The solid orange line and the solid lime curve
represent the $M-R$ relation in the zero temperature model for pure He and pure
Mg white dwarfs \citep{Hamada1961}, respectively. The solid red curve represents
white dwarfs in the BPS approach.

Note that in our modeling, the strange dwarfs are all covered by a
normal matter crust. The dashed curves in figure~\ref{fig:fig1}
are plotted for strange dwarfs with various crust-bottom
densities, with the density value marked near each curve (in units
of $\rm \, g\, cm^{-3} $). In fact, the radius of a strange dwarf
is mainly determined by the crust. The vertical bar
\textquotedblleft$ \mid $\textquotedblright ~marked with a letter
``b'' in each of the sequences represents the lightest object, and
the cross symbol \textquotedblleft $ \times $\textquotedblright
~marked with ``c'' indicates the most massive strange dwarf. The
cross symbol \textquotedblleft $ \times $\textquotedblright
~marked with ``d'' is the endpoint of the sequence in the case of
$ \rho_{\rm cb} = 4.3\times10^{11} \rm \, g\, cm^{-3} $. The SQM
core shrinks to almost zero at the end point of each curve.
Comparing with the white dwarf sequences, the strange dwarf
sequences are generally more compact. For example, for an object
with a typical mass of $0.8 \, M_{\odot}$, the radius of strange
dwarf is significantly less than that of white dwarf. It provides
us a useful clue to search for candidates of strange dwarfs.

\subsection{Comparison with Observations} \label{subsec:candidate}

\begin{figure}[h!]
	\centering
	\includegraphics[width=.75\textwidth]{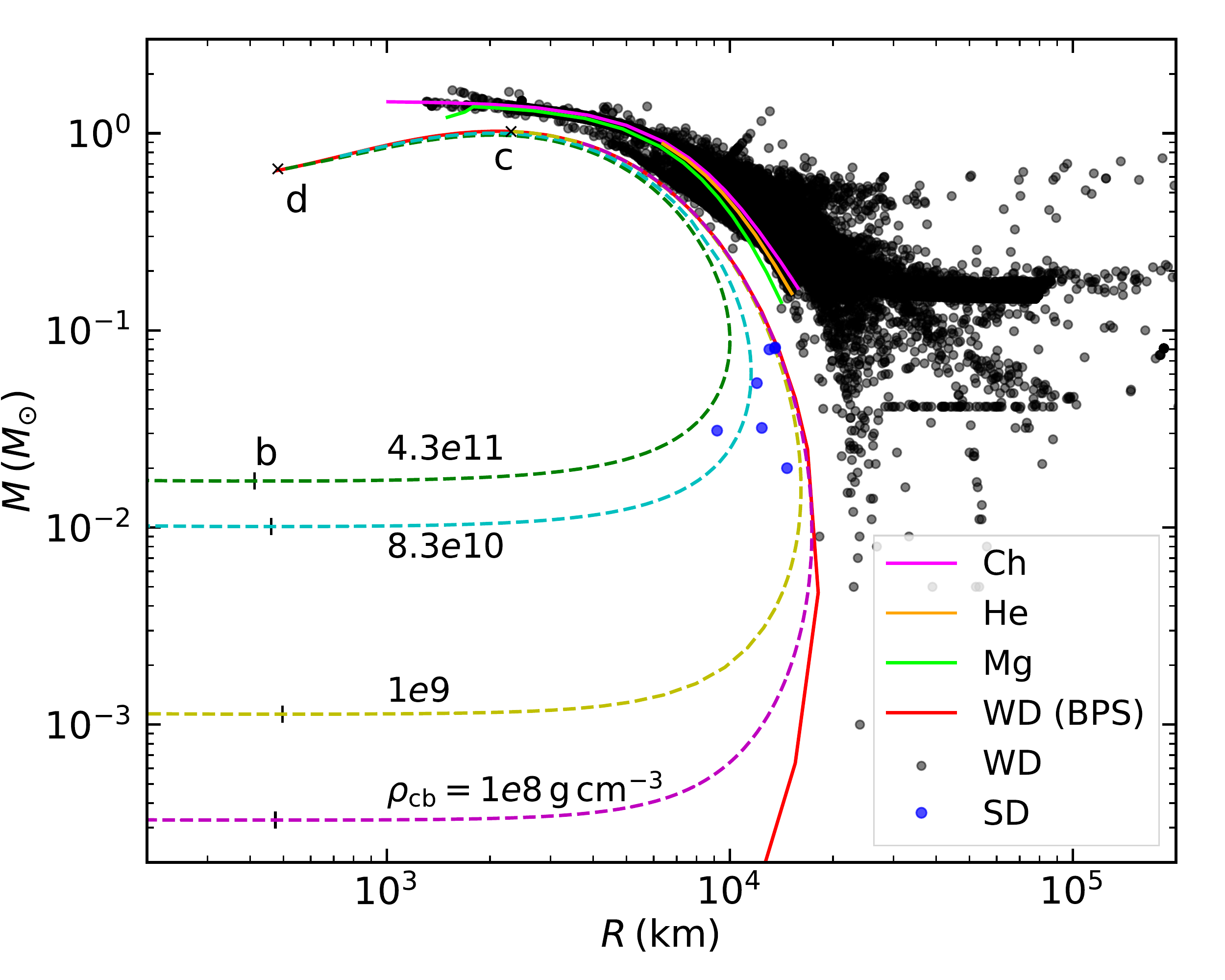}
	\caption{The mass-radius relations for the white dwarf and strange dwarf sequences.
		Different curves represent the mass-radius relations for compact dwarfs with different compositions.
		Observational data points corresponding to the 39041 white dwarfs in the
		MWDD database are also plotted.
		The black points represent normal white dwarfs and
		the blue points indicate candidates of strange dwarfs identified in
		this study. See the main text in Section~\ref{sec:selection} for more
		details of the curves and symbols.}
	\label{fig:fig1}
\end{figure}

In figure~\ref{fig:fig1}, we have also plotted the observational
data points which represents the 39041 white dwarfs in the MWDD
database. For the sake of clarity, we did not show the error bar
of each point. From this figure, we see that the majority of the
data points generally comply with the conventional white dwarf
theory \citep{Chandrasekhar1967,Chandrasekhar1935}. But there are
still some data points deviating from the conventional $ M-R $
relation. For example, some massive white dwarfs exceed the
Chandrasekhar mass limit of $1.44 \, M_{\odot}$. There are also
some white dwarfs having a large radius. These phenomena may be
caused by different composition inside the objects
\citep{Hamada1961,Panei2000}, or by strong magnetic field
\citep{Das2012,Ablimit2019a,Ablimit2019b,Gupta2020} and fast
rotation \citep{Das2013,Mukhopadhyay2016}.

\begin{figure}[t!]
	\centering
	\includegraphics[width=.75\textwidth]{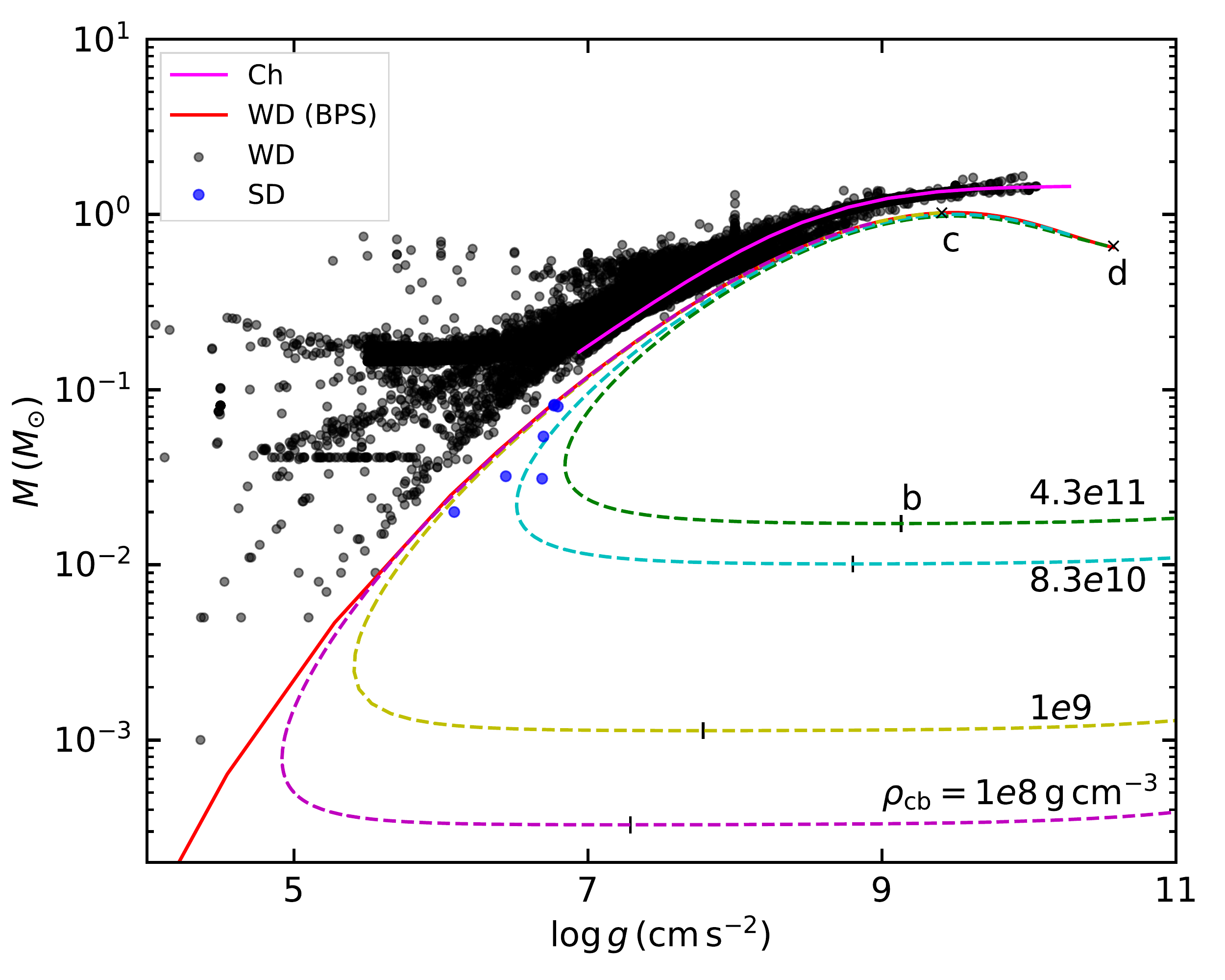}
	\caption{Mass versus the surface gravity for white dwarfs and strange dwarfs.
		Observational data points representing the 39041 white dwarfs in the
		MWDD database are also plotted.
		Line styles and symbols are the same as those
		in figure~\ref{fig:fig1}.}
	\label{fig:fig2}
\end{figure}

\begin{table}[h!]%[width=.98\textwidth,cols=6,pos=h!]
	\caption{Strange dwarf candidates in our sample.}\label{tab:table1}
	\begin{tabular}{llllll}
		\toprule
		%Col 1 & Col 2 & Col 3 & Col4 & Col5 & Col6\\
		MWDD ID & $ T_{\rm eff} $ & log\, g  & $ M $ &  $ R $ &   Ref. \\
		&  (K) & ($\rm cm \, s^{-2} $) & ($ M_\odot$) &  ($ \rm km $) &   \\
		\midrule
		LSPM J0815+1633     & 4655 $ \pm $ 35 & 6.772 $ \pm $ 0.076 & 0.082 $ \pm $ 0.031 & 13563.23 $ \pm $ 1024.76 &  \citep{Blouin2019}\\
		LP 240-30     & 4680 $ \pm $ 25 & 6.768 $ \pm $ 0.039 & 0.081 $ \pm $ 0.016 & 13542.5 $ \pm $ 626.6 &  \citep{Blouin2019}\\
		BD+20 5125B     & 4395 $ \pm $ 90 & 6.795 $ \pm $ 0.097 & 0.08 $ \pm $ 0.038 & 13046.72 $ \pm $ 1124.18 &  \citep{Blouin2019}\\
		LP 462-12     & 4800 $ \pm $ 20 & 6.697 $ \pm $ 0.054 & 0.054 $ \pm $ 0.024 & 11999.23 $ \pm $ 1552.78 &  \citep{Blouin2019}\\
		WD J1257+5428     & 7485 $ \pm $ 85 & 6.441 $ \pm $ 0.068 & 0.032 $ \pm $ 0.03 & 12403.13 $ \pm $ 3561.25 &  \citep{Blouin2019}\\
		2MASS J13453297+4200437     & 4270 $ \pm $ 75 & 6.688 $ \pm $ 0.086 & 0.031 $ \pm $ 0.04 & 9186.23 $ \pm $ 3405.51 &  \citep{Blouin2019}\\
		SDSS J085557.46+053524.5   & 10670 $ \pm $ 1677 & 6.09 $ \pm $ 1.078 & 0.02 $ \pm $ 0.245 & 14688.29 $ \pm $ 767.07 &  \citep{Rebassa-Mansergas2016}\\
		\bottomrule
	\end{tabular}
\end{table}

However, among the observed 39041 ``white dwarfs'', we notice that
seven stars appear to be quite special. They obviously deviate
from the mass-radius curves for white dwarfs, but well match the
relation of strange dwarfs. Their masses are in the range of $\sim
0.02 $ --  $0.12 \, M_{\odot}$ and their radii are $\sim$ 9,000 --
15,000 km. In other words, they are too compact to be normal white
dwarfs. We argue that these seven objects are good candidates for
strange dwarfs. In figure~\ref{fig:fig1}, we have specially marked
these strange dwarf candidates with blue color. Some key
parameters of them are listed in table~\ref{tab:table1}. Note that
among these seven dwarfs, the parameters of six objects (LSPM
J0815+1633, LP 240-30, BD+20 5125B, LP 462-12, WD J1257+5428 and
2MASS J13453297+4200437) were derived based on the spectroscopic
data and the Gaia observations \citep{Blouin2019}. In their
calculations, H/He composition of the atmosphere has been taken
into account (see table 3 in ref. \citep{Blouin2019}). Note that
they also pointed out that no evidence for binary interaction is
observed for most of these objects.

Figure~\ref{fig:fig2} further illustrates the mass as a function
of the surface gravity. The currently available 39041 white dwarfs
in the MWDD database are also plotted for comparison. Again, we
see that the seven candidates deviate from the white dwarf
sequences, but are better matched by the strange dwarf sequences.

\begin{figure}[h!]
	\centering
	\includegraphics[width=.75\textwidth]{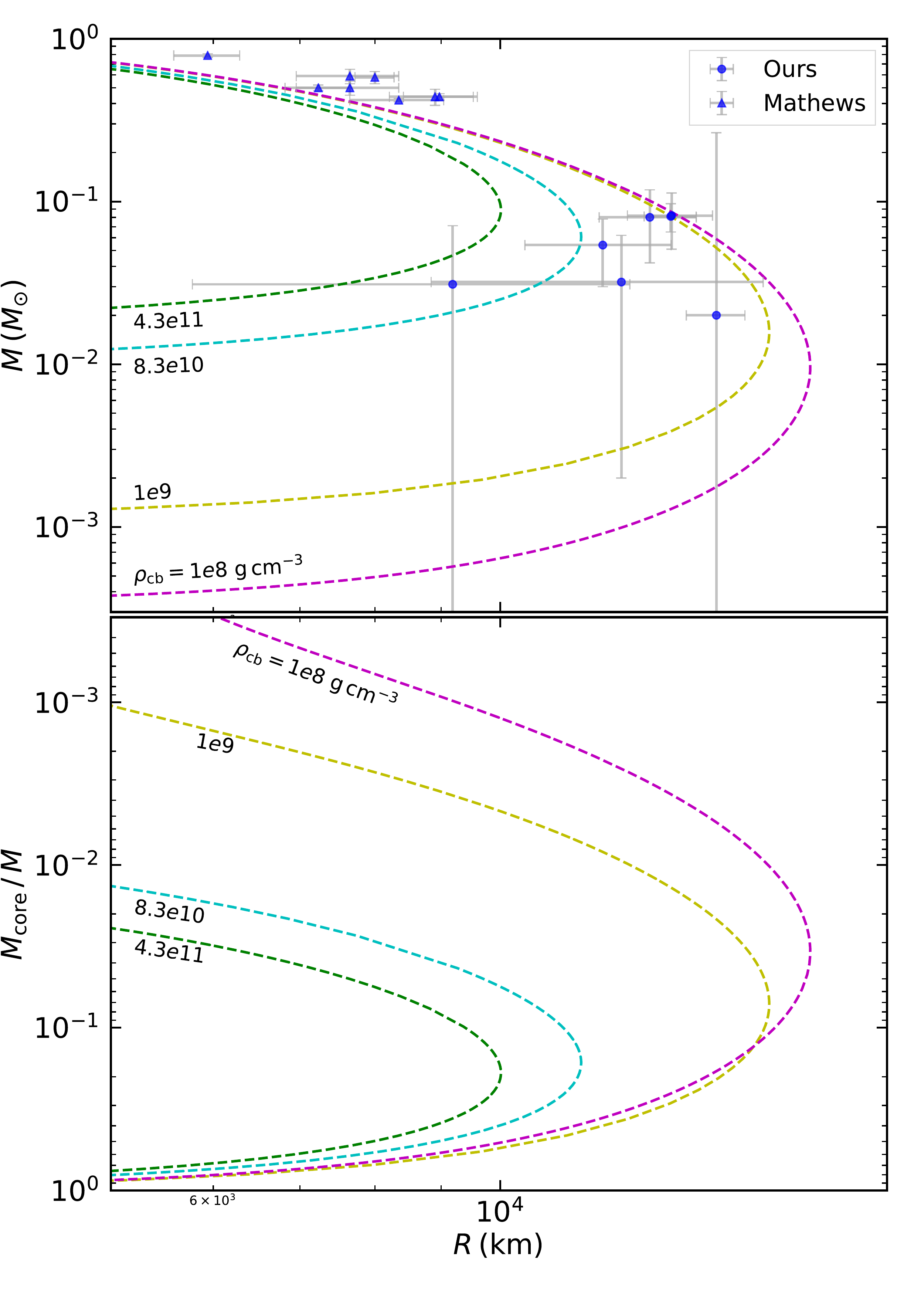}
	\caption{A zoomed-in plot of the mass-radius relation of strange dwarfs (upper panel),
		and the SQM fraction versus the total radius (lower panel).
		In the upper panel, the blue circles represent the strange dwarf candidates
		identified in this study, and the blue triangles correspond to the candidates
		suggested in \citep{Mathews2006}.
		In both panels, the curves are plotted by assuming different crust bottom densities (marked
		near each curve, in units of $\rm g \, cm^{-3}$).}
	\label{fig:fig3}
\end{figure}

\section{Comparison with previous studies} \label{sec:comare}

It is interesting to note that Mathews et al \citep{Mathews2006} have also suggested
a few compact objects as candidates of strange dwarfs. To give a direct comparison
with their results, we present a zoomed-in plot of the mass-radius
relation for strange dwarfs in figure~\ref{fig:fig3}. From this figure,
we see that candidates of \citep{Mathews2006} are mainly in the mass range
of 0.4 -- 0.8 $M_\odot$ and the radii are 6,000 -- 9,000 km. In such a
high-mass region, the difference between strange dwarfs and white dwarfs
on the mass-radius plot is actually very small. In fact,
all the candidates of \citep{Mathews2006} lie only slightly above the
dashed curves, which means they still seem to be not compact enough.
On the contrary, our
candidates are in the low-mass region, with masses between 0.02 -- 0.12
$M_\odot$ and radii between 9,000 -- 15,000 km. In this region, the
difference between strange dwarfs and white dwarfs is quite significant.
So, it is a more appropriate region for discriminating between these two
kinds of dwarfs. From figure~\ref{fig:fig3}, we clearly see that our
candidates can only be fitted by the strange dwarf curves. They obviously
deviate from the white dwarf sequences even when the error bars are
considered (c.f. figure~\ref{fig:fig1}).
Figure~\ref{fig:fig3} also shows
that these strange dwarfs generally should have a crust bottom density
of $\rm 1.0\times10^{9}\,g\,cm^{-3} \leq \rho_{\rm cb}
\leq \rm 1.0\times10^{11}\,g\,cm^{-3}$, which is significantly smaller than
the neutron drip density ($\rho_{\rm drip}$).
From figure~\ref{fig:fig3}, we argue that our sample are more credible
strange dwarf candidates.

The lower panel of figure~\ref{fig:fig3} shows the SQM fraction versus the
stellar radius of strange dwarfs. Here, the SQM fraction is defined as the
mass percentage of the SQM core with respect to the whole dwarf star
($ M_{\rm core}/M $). We see that the SQM fraction of low-mass strange dwarfs are
higher than that of high-mass ones on each curve (see Section \ref{subsec:m-r}
for the variation of $ M_{\rm core} $). The SQM fraction is about
$ 3.58 \% $ ($ M = 9.1\times10^{-3} \, M_{\odot} $, $ M_{\rm core} = 3.26\times10^{-4} \, M_{\odot} $) for the largest
radius object in the case of $ \rho_{\rm cb} = 10^{8} \rm \, g\, cm^{-3}$,
and is about $ 19.2 \% $ ($ M = 8.75\times10^{-2} \, M_{\odot} $, $ M_{\rm core} = 1.68\times10^{-2} \, M_{\odot} $)
for the largest radius object in the case of $ \rho_{\rm cb} = 4.3\times10^{11} \rm \, g\, cm^{-3}$.
Since the candidates of \citep{Mathews2006} are generally more massive, their
SQM fractions are typically much smaller than 0.001.
However, the SQM fraction of our candidates are larger than 0.001, which means the SQM
core plays a more important role inside the star so that its observational effects
could be more significant. For example, it may affect the cooling process of the
star (see discussion in the next section).
In fact, the strange dwarf candidates in our sample generally have
a core mass of $ M_{\rm core} \sim 10^{-4} $ -- $ 10^{-2}\, M_{\odot}$.
The relatively larger SQM fraction of less massive strange dwarfs further
supports our scheme of trying to search for SQM objects among smaller white dwarfs.

\section{Discussion and Conclusions} \label{sec:conclusion}

In this study, we have tried to search for strange quark dwarf candidates
among the observed white dwarfs. The selection of candidates
is conducted by comparing the mass and radius of observed dwarfs with
the theoretical mass-radius relation of strange dwarfs.
It is found that the masses and radii of seven objects are consistent with
that of strange dwarfs. The mass of the SQM core of these candidates ranges
in $\sim 10^{-4} $ -- $ 10^{-2} \, M_{\odot} $. The seven objects are
LSPM J0815+1633, LP 240-30, BD+20 5125B, LP 462-12,
WD J1257+5428, 2MASS J13453297+4200437, and SDSS J085557.46+053524.5.
Note that although SDSS J085557.46+053524.5 is in a binary system \citep{Rebassa-Mansergas2016},
the other six are single according to currently available observations~\citep{Blouin2019}.

How strange dwarfs are produced is an interesting question. It has been
suggested that strange dwarfs may be formed either from the capture of
strange nuggets by stellar objects (main-sequence stars, brown dwarfs, etc.),
or from the growth of strange clumps by accreting from ambient medium.
Below are several detailed formation channels for strange dwarfs. Firstly, the
Universe once went through a quark era in its expansion history according to
the big bang theory. At that time, a large amount of SQM clumps
may be formed and survive up to
now \citep{Cottingham1994}. They can be captured by small
main-sequence stars and fall into the center of the star,
and convert the normal matter into SQM \citep{Glendenning1995}.
Secondly, explosive events such as supernovae, merging of two compact stars,
phase transition of massive NSs may be responsible for the formation
of SSs, and a large amount of
SQM nuggets may be ejected during these processes.
The flux of SQM nuggets produced in this way
in a typical galaxy is estimated as $ \sim 0.1 \,\rm cm^{-2}~s^{-1} $
\citep{Madsen1988,Glendenning1990,Glendenning1992}.
These SQM nuggets can contaminate surrounding objects,
and convert them into strange dwarfs \citep{Olinto1987}.
Thirdly, a newborn SS is quite hot and highly turbulent.
It may directly eject a large clump of SQM due to
the joint effect of fast-spinning and turbulence
\citep{Xu2003,Horvath2012,Xu2006}. The SQM clump then can accrete ambient matter
throughout its life and evolve to its current form.

A remarkable feature of the strange dwarf candidates in our sample is
that their temperatures and masses are very low (see table~\ref{tab:table1}).
Generally, their temperatures are cooler than 13,000 K and their masses
are in the range of $ 0.02 $ -- $ 0.12 \, M_{\odot} $.
Many authors have discussed this feature in the framework of white dwarfs.
Low-mass white dwarfs are likely formed in
binary systems because it is impossible for a star with a mass below
$ 0.45 \, M_{\odot} $ to become a white dwarf through single-star evolution within
the age of the universe \citep{Liebert2005,Rebassa-Mansergas2011}. On the contrary,
in a binary system, the progenitor star can lose mass through binary interactions
\citep{Benvenuto2005,Ablimit2019b,Marsh1995,Sun2018}.
This opinion is supported by the observations of several extremely low-mass white dwarfs
in compact binaries involving NSs/pulsars \citep{Benvenuto2005,Marsh1995, Driebe1998,
	Lorimer2008} and WDs \citep{Brown2016,Brown2020ApJ}.
However, there is no evidence for binarity for most of our candidates
according to present observations. As for the low temperature,
Blouin et al. \citep{Blouin2019} suggested that these ultra-cool objects
may be polluted by rocky debris, but they did not provide any information about
the origin of these rocky debris around such low-mass objects.
%because this process is independent of evolution.
In this study, we have identified these objects as strange dwarf candidates.
The low temperature might be self-consistently explained as due to the effect
of the SQM core inside them. In fact, SQM objects generally cool faster so that
their surface temperature is usually lower than their hadronic matter
counterparts \citep{Pizzochero1991}.
It is interesting to note that more candidates of small SQM stars up to planetary SQM
objects have been suggested recently by \citet{Huang2017,Kuerban2020,Kuerban2019}.
Especially, the planet PSR 1719-14 b is generally regarded as an appropriate candidate because
it has a density at least larger than 23 $\rm g \,cm^{-3}$ \citep{Horvath2012,Huang2017,Kuerban2020,Kuerban2019,Bailes2011}.
In the future, these objects deserve being extensively examined by using large optical/infrared telescopes
and radio facilities.

\bigskip
%\acknowledgments
\noindent \textbf{Acknowledgments}: We would like to thank the anonymous
referees for helpful suggestions that led to significant
improvement of our study. This work is supported by the special
research assistance project of the Chinese Academy of Sciences
(CAS), the National Key R\&D Program of China (grant No.
2021YFA0718500), the CAS ``Light of West China'' Program (grant
No. 2018-XBQNXZ-B-025),  the National SKA Program of China No.
2020SKA0120300, the National Natural Science Foundation of China
(grant Nos. 11873030, 12041306, U1938201, 11535005, 11690030,
11903019, 12033001, 12041304, and 11873040), the Operation,
Maintenance and Upgrading Fund for Astronomical Telescopes and
Facility Instruments, budgeted from the Ministry of Finance of
China (MOF) and administrated by the CAS, and by the science
research grants from the China Manned Space Project with grant No.
CMS-CSST-2021-B11. This work has made use of the Montreal White
Dwarf Database.



%% References

%\bigskip
%\noindent \textbf{References}
\bibliographystyle{model1-num-names}
\bibliography{cas-refs}



%\begin{thebibliography}{00}


%\end{thebibliography}

\end{document}